# SUPER-COMPACT SLED SYSTEM USED IN THE LCLS DIAGNOSTIC SYSTEM*

J. W. Wang[#], S.G. Tantawi, X. Chen, SLAC, Menlo Park, CA 94025, USA

*Abstract*

At SLAC, we have designed and installed an X-band radio-frequency transverse deflector system at the LCLS for measurement of the time-resolved lasing effects on the electron beam and extraction of the temporal profile of the pulses in routine operations.

We have designed an X-Band SLED system capable to augment the available klystron power and to double the kick.

## INTRODUCTION

As shown in Figure 1, Radiofrequency (RF) transverse deflectors are used in conjunction with an electron beam energy spectrometer to measure the electron beam longitudinal (time-energy) phase space downstream of the FEL undulators. Since the lasing process in an FEL induces both electron energy loss and energy spread growth, the output electron longitudinal phase space will include a 'footprint' left by the emitted X-rays [1].
Enabled by well-developed RF technology, this diagnostic method is simple, single shot and non-invasive to FEL operation. Furthermore, it covers a wide dynamic range and works for both soft and hard X-ray energies without requiring any configuration change.

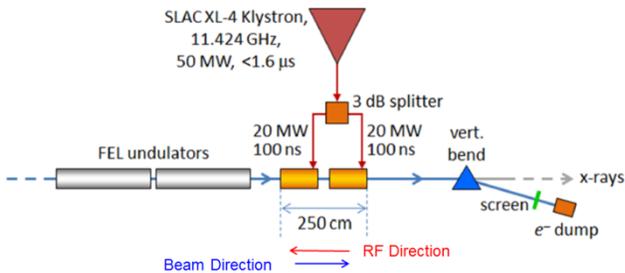

Figure 1: Layout of deflector RF system downstream of the LCLS undulators.

- The kick for a 1m Section is $5.46 (P_{in})^{1/2}$ MV, where $P_{in}$ is the Peak RF Power in MW. The maximum kick of little more than 40 MV was obtained from an old Klystron with peak power of only 35 MW. In order to reach higher resolution, we have been designing a brand new SLED system to double the kick to more than 80MV. This paper describes its principle and technical advances and challenges.

## SLED ENERGY GAIN

In June 1974, forty years ago, a clever invention named "SLED" was announced at SLAC [2]. The increase of the peak RF power is in exchange for the RF pulse length reduction by a passive technique called "pulse compression". As shown in Figure 2, the first part of klystron pulse was stored in two low-loss cylindrical tuned cavities installed downstream of the klystron. For the remaining part of the pulse (usually to be the filling time of the feeding accelerator structure), the phase of the klystron was reversed by 180 electrical degrees, and the sum of the power now being discharged by the cavities plus the direct klystron power resulted in a net power gain. The key components of a SLED system include a 3db coupler with two 90° apart divided power ports and high Q energy storage cavities.

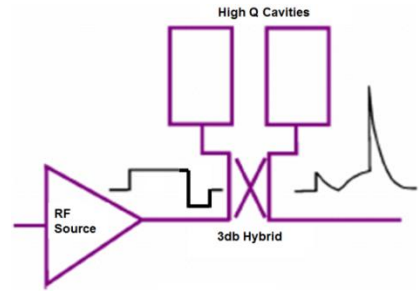

Figure 2: Basic SLED working scheme.

Our system can be described as followings: over-coupling coefficient $\beta=Q_0/Q_e$, where $Q_o=10^5$ is unloaded Q and $Q_e$ and $Q_L$ are external and loaded Q respectively. The cavity filling time Tc is $2Q_L/\omega=2Q_0/\omega(1+\beta)$, the coupling coefficient β will be optimized later. The deflector is 1.0 m long, constant impedance structure with transverse impedance of 41.9 MΩ/m, filling time $T_f$=106 ns (group velocity of -3.165 % speed of light) and attenuation factor τ=0.62 Neper. Let us assume the total RF pulse is 1.5 μs and last 106 ns has been flipped phase with 180°. Figure 3 shows loaded voltage waveform after SLED.

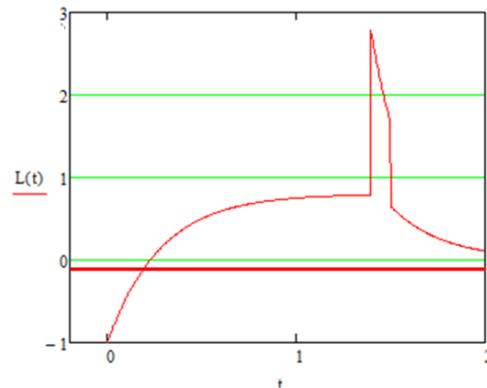

Figure 3: Loaded SLED voltage waveform with horizontal coordinate of time in μs and vertical coordinate in none SLEDed input voltage as unit.

___________________________________________

* Work supported by Department of Energy contract DE–AC03–76SF00515.

# jywap@slac.stanford.edu

With this type of loaded voltage to the backward wave deflector, we can calculate total kick voltage for a beam bunch in any injection time. The calculated result is shown in Figure 4.

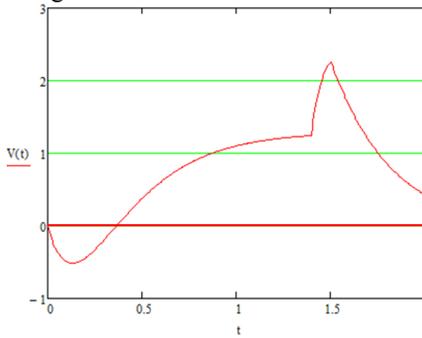

Figure 4: Calculated total kick voltage as a function of beam injection time with the vertical coordinate in none SLEDed input voltage as unit.

It is interesting to study the kick voltage along the deflector for existing constant impedance structure in comparison with a constant gradient structure. Figure 5 shows the kick voltage is much flat with SLEDed pulse for constant impedance structure.

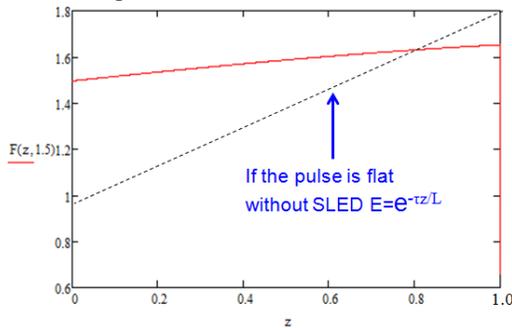

Figure 5: Kick voltage along the deflector structure for constant impedance structure (red) and constant gradient structure (black).

Finally, we need to optimize the SLED system by calculating its total gain for various coupling coefficient of high Q cavities. Figure 6 shows the highest gain for Q~10 could be larger than 2 for the over-coupling coefficient β around 9-10.

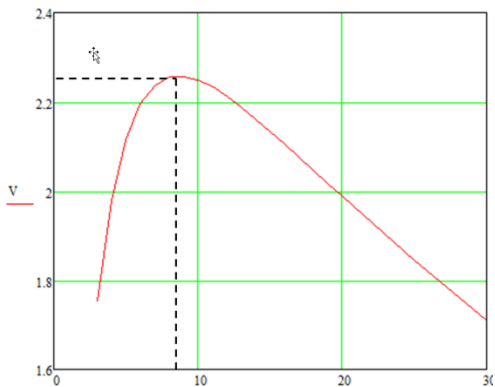

Figure 6: Total SLED gain as function of coupling coefficient β for high Q cavities.

## DUAL-MODE CIRCULAR POLARIZER

Having all the basic functions of a 3db coupler, a much compact and elegant dual-mode circular polarizer was developed to convert the $TE_{01}$ mode in a rectangular waveguide into two polarized $TE_{11}$ modes in quadrature into a circular waveguide [3] (see Figure 7).

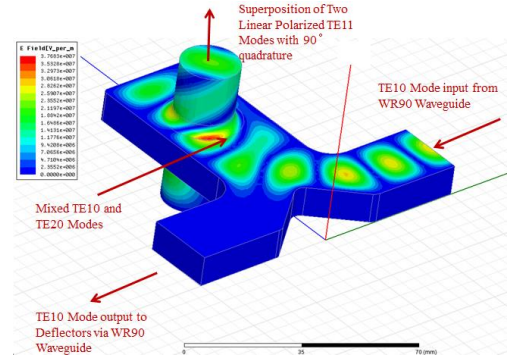

Figure 7: Schematic view of the dual-mode polarizer.

As shown in the figure, the $TE_{01}$ mode converts to both $TE_{01}$ and $TE_{02}$ modes in a wide waveguide region, and their magnetic field components would couple to two perpendicular polarized $TE_{11}$ modes in the circular waveguide. If the geometry is properly chosen, the two polarized mode can be adjusted in phase quadrature.

Naturally, this unique device can feed a single sphere cavity with two respective sphere modes as SLED storage cavity as shown in Figure 8.

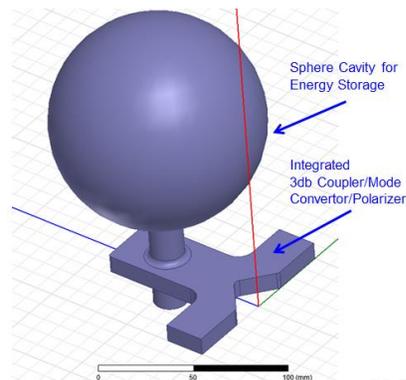

Figure 8: Schematic view of the new SLED system.

## SPHERE CAVITY MODES

The electrical vector potential F for all modes in a sphere with radius r=a can be described as:

$$(F_r)_{mnp} = \hat{J}_n\left(u_{np}\frac{r}{a}\right) P_n^m(\cos\vartheta) \begin{Bmatrix} \cos m\varphi \\ \sin m\varphi \end{Bmatrix}$$

where $\hat{J}_n$ is sphere Bessel Function and $P_n^m$ is associated Legendre Polynomials with m≤n.

For TE modes, the $E_\varphi = H_\theta = 0$ at surface r=a. It means $\hat{J}_n(u_{np})=0$. The mathematics tables show the values of $u_{np}$ for all the lower order modes. Therefore, the sphere radius can be calculated using wave propagation constant k and value of $u_{np}$.

$$a = \frac{u_{np}}{k} = \frac{c \times u_{np}}{k} = 0.41767 u_{np}\ (cm)$$

Sphere radius is independent with mode index m, there are numerous degeneracies because $\hat{J}_n(u_{np})$ is independent with m.

Practically, we have chosen $TE_{m14}$ modes. There are three possible modes:

$$(F_r)_{014} = \hat{J}_1\left(14.066\frac{r}{a}\right)\cos\vartheta$$
$$(F_r)_{114} = \hat{J}_1\left(14.066\frac{r}{a}\right)\sin\vartheta\cos\vartheta$$
$$(F_r)_{114} = \hat{J}_1\left(14.066\frac{r}{a}\right)\sin\vartheta\sin\vartheta$$

For perfect sphere cavity, these three modes have the same mode patterns except that they are rotated 90° in space from each other, which can be seen in following figures.

In reality, they can be slightly distinguished in frequencies due to the perturbation from the different coupling in the coupler port. The $TE_{014}$ mode is higher in frequency and very weekly under-coupled due to the feeding orientation.

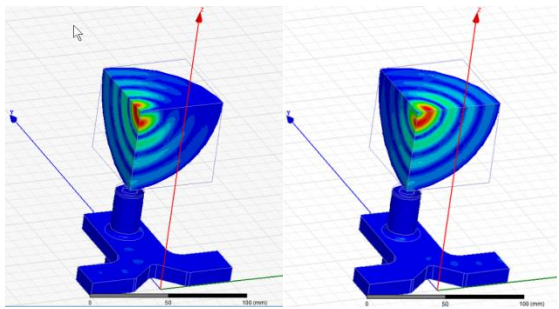

Figure 1: TwError! Hyperlink reference not valid.o differently irritated $TE_{114}$ modes as SLED working modes.

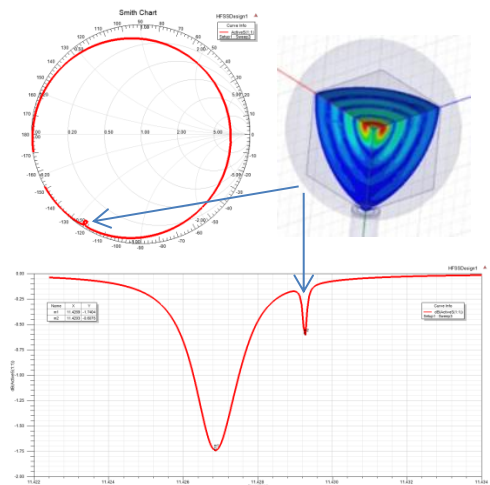

Figure 1: From the HFSS simulation, the separation of $TE_{114}$ modes and $TE_{014}$ mode is about 2.3 MHz. $TE_{014}$ mode is under-coupled with β~0.035 and $TE_{114}$ modes is optimized coupled with β~10.

Another interesting property is the sole dependence of $Q_0$ on the sphere radius without depending on the mode types. The quality factor $Q_0$ for TE Modes is

$$Q_0 = \frac{a}{\delta}$$

where δ is the skin depth (for Copper 0.61μm).

For example, in our case of the $TE_{114}$ mode at 11424 MHz, the sphere radius a~5.8749 cm, $Q_0=0.963\times10^5$, the optimized SLED gain is 2.2 for β= 8-9.

## INTEGRATED NEW SLED DESIGN

Based on above abatements, we have integrated the new SLED system design as shown in Figure 11.

The vacuum pump will be installed at the bottom match cylinder of the dual mode polarizer.

Tuning will be performed by combination of matching a ridge in the sphere equator and push-pull tuning on the top region of the sphere.

The detuning for non-SLEDed mode will be done by inserting a tungsten middle into the sphere cavity.

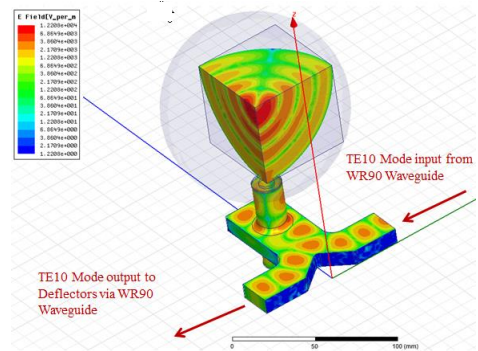

Figure 11: Schematic view of the wave propagation in the SLED system.

## OUTLOOK OF NEW SLED SYSTEM

After extensive studies for maximum fields, cooling requirement and manufacturability issues, we are in the mechanical design phase for the complete system. The preliminary schedule is to start machining in sometime October, 201, to have a microwave characterization by the end of 2014 and high power test in 2015.

Because of its super characteristics as well the compactivity and simplicity of this new SLED system, it can be easily to be applied to C-Band, S-Band and more other cases of frequencies.

Furthermore, because of the feature with abundant high Q modes in the neighbourhood of SLED working frequency in a sphere resonant cavity, by properly choosing and tuning them, we predict a bright future to create a rather flat SLED pulses.